\documentclass[useAMS,usenatbib]{mn2e}

\usepackage[a4paper,centering, totalwidth=520pt, totalheight=700pt]{geometry}
\usepackage{aas_macros}
\usepackage{natbib}
\usepackage{amsfonts}
\usepackage{amsmath}
\usepackage{amssymb}
\usepackage{graphicx}
\usepackage{color}

\usepackage{microtype}

\newcommand{\Mh}{\ensuremath{h^{-1}M_{\odot}}}

\newcommand{\avg}[1]{\ensuremath{\left\langle \,#1\, \right\rangle}}

\newcommand{\der}{\ensuremath{{\rm d}}}

\newcommand{\delc}{\ensuremath{\delta_{\rm c}}}
\newcommand{\delo}{\ensuremath{\delta_{0}}}
\newcommand{\sig}{\ensuremath{\sigma}}

\newcommand{\erfc}[1]{\ensuremath{{\rm erfc}\left(#1\right)}}
\newcommand{\erfcinv}[1]{\ensuremath{{\rm erfc}^{-1}\left(#1\right)}}

\newcommand{\eqn}[1]{equation~\eqref{#1}}

\newcommand{\be}{\begin{equation}}
\newcommand{\ee}{\end{equation}}

\newcommand{\HI}{\textsc{Hi}}
\newcommand{\HII}{\textsc{Hii}}

\begin{document}
\pagerange{\pageref{firstpage}--\pageref{lastpage}} 
\date{draft}

\title[Photon Conserving Bubbles]
{Photon Number Conserving Models of {\sc Hii} Bubbles during Reionization}
\author[Paranjape et al.]
{Aseem Paranjape$^1$\thanks{E-mail: aseem@iucaa.in},~
T. Roy Choudhury$^2$\thanks{E-mail: tirth@ncra.tifr.res.in}~
and
Hamsa Padmanabhan$^{1,3}$\thanks{E-mail: hamsa.padmanabhan@phys.ethz.ch}
\\
$^1$Inter-University Center for Astronomy \& Astrophysics, Post Bag 4, Ganeshkhind, Pune 411007, India.\\
$^2$National Centre for Radio Astrophysics, TIFR, Post Bag 3, Ganeshkhind, Pune 411007, India.\\
$^3$Institute for Astronomy, ETH Zurich, Wolfgang-Pauli-Strasse 27, CH 8093 Zurich, Switzerland.
} 

\maketitle
\label{firstpage}

\begin{abstract}
\noindent
Traditional excursion set based models of \textsc{Hii} bubble growth during the epoch of reionization are known to violate photon number conservation, in the sense that the mass fraction in ionized bubbles in these models does not equal the ratio of the number of ionizing photons produced by sources and the number of hydrogen atoms in the intergalactic medium. 
E.g., for a Planck13 cosmology with electron scattering optical depth $\tau\simeq0.066$, the discrepancy is $\sim15$ per cent for $x_{\textsc{Hii}}=0.1$ and $\sim5$ per cent for $x_{\textsc{Hii}}=0.5$. 
We demonstrate that this problem arises from a fundamental conceptual shortcoming of the excursion set approach (already recognised in the literature on this formalism) which only tracks \emph{average} mass fractions instead of the exact, stochastic source counts. 
With this insight, we build an approximately photon number conserving Monte Carlo model of bubble growth based on partitioning regions of dark matter into halos. 
Our model, which is formally valid for white noise initial conditions (ICs), shows dramatic improvements in photon number conservation, as well as substantial differences in the bubble size distribution, as compared to traditional models. 
We explore the trends obtained on applying our algorithm to more realistic ICs, finding that these improvements are robust to changes in the ICs. 
Since currently popular semi-numerical schemes of bubble growth \emph{also} violate photon number conservation, we argue that it will be worthwhile to pursue new, explicitly photon number conserving approaches. 
Along the way, we clarify some misconceptions regarding this problem that have appeared in the literature. 
\end{abstract}

\begin{keywords}
dark ages, reionization, first stars -- intergalactic medium -- cosmology: theory -- large-scale structure of Universe
\end{keywords}

\section{Introduction}
\label{sec:intro}
\noindent
The 21cm signal arising from cosmological neutral hydrogen (\textsc{Hi}) during the epoch of reionization ($6 \lesssim z \lesssim 15$) is of great importance for next generation astrophysical and cosmological studies, primarily due to its close connection with the formation of the first stars \citep[for reviews, see][]{fob06,mw10,pl12}. Detecting this signal -- which is expected to be buried under astrophysical foregrounds that are orders of magnitude stronger -- would herald a new phase in our understanding of the high redshift Universe. This signal is being targeted by various current (GMRT\footnote{http://gmrt.ncra.tifr.res.in/}, PAPER\footnote{http://eor.berkeley.edu/}, LOFAR\footnote{http://www.lofar.org/}, MWA\footnote{http://www.mwatelescope.org/}) and more sensitive upcoming instruments (SKA\footnote{https://www.skatelescope.org/}, HERA\footnote{http://reionization.org/}) by measuring the fluctuations in the \textsc{Hi} density. This huge observational effort is being complemented by remarkable progress in making accurate and robust theoretical predictions of the observable signal; building this theoretical understanding is essential due to the expected weakness of the signal and its rich potential in constraining both astrophysics and cosmology in the high redshift Universe.

In standard scenarios where galaxies are the dominant sources of ionizing photons, reionization proceeds through the growth and overlap of ionized ``bubbles'' around galaxies. The distribution of these bubbles will lead to fluctuations in the \HI~density field, which essentially forms the cosmological signal of interest to the telescopes. Hence, understanding the distribution of these ionized bubbles forms the first step in modelling the observable signal. It turns out that modelling these bubbles is not straightforward mainly due to uncertainties in various complex physical processes at high redshifts. The most rigorous way of accounting for all the complexities is possibly through numerical simulations which include detailed radiative transfer calculations \citep{gnedin00,cfw03,mfc03,iliev+06a,iliev+06b,mellema+06,mcquinn+07,stc08,tcl08,kg15}. Unfortunately, the dynamic range required in these simulations to probe the relevant processes is still beyond the computing power available at present. In addition, since these simulations are often very resource intensive, they are not very suitable for probing the range of uncertain parameters. It is thus useful to complement them with analytical models which are fast and can give reasonably accurate results with minimum number of approximations.

A fast and convenient method for modelling the distribution of bubbles is to use a formalism based on the excursion set approach \citep{bcek91}, as was pioneered by \citet*[][hereafter, FZH04]{fzh04}. The algorithm is based on counting the number of ionizing photons produced by sources in a region and comparing with the number of hydrogen atoms. The condition for the region to be ionized is that the number of ionizing photons in it exceeds the number of hydrogen atoms (adjusted for the number of recombinations, assuming it to be uniform in the medium). This method can naturally account for the overlap of ionized regions by finding the largest region for which the ionization condition is satisfied. The outputs of the algorithm include the size distribution of unique, non-overlapping ionized bubbles as well as the integral of this distribution, which is the mass fraction contained in such bubbles.

This excursion set based algorithm of FZH04 (which we discuss in detail below) subsequently formed the basis for a wide range of semi-numerical models \citep{mf07,santos+08,zahn+07,chr09,mfc11-21cmfast,lofs16}  for ionized bubbles, many of which have been found to give a good match with radiative transfer simulations \citep{majumdar+14}. However, a closer inspection of the formalism shows that it violates the conservation of the number of photons, in the sense that the mass fraction in ionized bubbles as returned by the algorithm does not equal the ratio of the number of ionizing photons produced by all the sources (compensated for recombinations in the medium) and the number of hydrogen atoms in the intergalactic medium, implying that photons are apparently being either gained or lost in the process\footnote{As we will discuss below, in the FZH04 model the mass in ionized bubbles is always less than the ratio of ionizing photons to hydrogen atoms, thus implying loss in the number of photons.}. Although this feature of the algorithm has been noticed earlier in the literature \citep[see, e.g.,][]{zahn+07}, we argue below that no satisfactory solution has been proposed to date.

The main aim of this work is to critically examine the origin of the non-conservation and subsequently develop a model which is free from this drawback. As we will demonstrate, it is indeed possible to construct a rigorous  photon-conserving theoretical model, at least for the restricted case where the initial conditions (ICs) for the density fluctuations have a white noise power spectrum. Although these ICs are not realistic, the analysis is nevertheless important because it provides a clear direction towards understanding the origin of the non-conservation in excursion set models of \textsc{Hii} bubbles. Along the way we will also clarify certain misconceptions in the literature regarding the origin and nature of this shortcoming of standard excursion set models, and show that the non-conservation extends to excursion-set based semi-numerical schemes as well, making this an important problem to tackle.

The plan of the paper is as follows. In section~\ref{sec:critique} we first argue that any traditional excursion set model of bubble size distributions will not conserve photon number in general, and then demonstrate this explicitly for the FZH04 model. In section~\ref{sec:numconsbub}, we use the insights gained by this exercise to construct a bubble model which \emph{does} conserve photon number if the ICs have a white noise power spectrum, and discuss generalisations of this model to more realistic initial power spectra. We conclude in section~\ref{sec:conclude} with a brief discussion of future directions. 
The Appendices provide technical details of some of the results used in the main text.
For several comparisons we will use a flat $\Lambda$-cold dark matter ($\Lambda$CDM) cosmology with parameters $\Omega_{\rm m}=0.315$, $\Omega_{\rm b} = 0.0487$, $h=0.673$, $\sigma_8=0.83$ and $n_{\rm s}=0.96$; we will refer to this as the `Planck13' cosmology \citep{Planck13-XVI-cosmoparam}.

\section{Critique of excursion set models of HII bubble growth}
\label{sec:critique}
\noindent
The starting point in excursion set models of the growth of \textsc{Hii} bubbles (FZH04) is to define an \textsc{Hii} bubble as a region that has just enough sources of ionizing photons to be completely ionized. Ionizing sources are assumed to reside in all dark matter halos more massive than a threshold mass $m_{\rm min}$, and a source in a halo of mass $m$ is assumed to ionize an \textsc{Hi} mass $\zeta m$, where $\zeta$ is the number of ionizing photons per baryon multiplied by the baryonic fraction\footnote{The parameter $\zeta$, in principle, includes the effect of recombinations in the IGM provided the number of recombinations is uniform. We ignore effects of non-uniform recombinations and self-shielding in high density regions.}. For simplicity, in the following we will assume $\zeta$ to be constant, although all our results can be easily generalised to a mass- and/or time-dependent $\zeta$. An \textsc{Hii} bubble is then a region of mass $M_b$ which satisfies the condition that it is the \emph{largest} region for which $M_b = \sum_h\zeta m_h$ where the sum runs over all sources inside this region, with halo masses $m_h$. In other words, a region of mass $M_b$ is a bubble if the mass fraction in sources in this region equals $1/\zeta$ and the region is not embedded in any larger region satisfying this condition.

The excursion set framework provides a natural formal setting for calculating the mass distribution of such regions. This is because it is possible in this framework to calculate the \emph{average} mass fraction $f(m|\delo,M_0)$ in halos (i.e., sources) of mass $m$ in regions of fixed mass $M_0$ and overdensity $\delo$, integrating which leads to the average mass fraction in all sources $f(>m_{\rm min}|\delo,M_0)$. FZH04 advocated using this quantity to determine the bubble mass distribution by calculating the distribution of masses $M_0 = M_b$ at which the collapse fraction first crosses the ``barrier'' $1/\zeta$ as the mass $M_0$ is decreased from large values. Since $\delo$ is a stochastic function of $M_0$ for a given set of ICs, this problem is very similar to the one solved by \citet{bcek91} in determining $f(m|\delo,M_0)$ (there, the barrier was the critical density for spherical collapse $\delta_{\rm c}$). Most analytical and semi-numerical schemes for determining the bubble distribution \citep{mf07,santos+08,zahn+07,chr09,mfc11-21cmfast,pc14,lofs16} are based on variations of this basic framework, with differences largely in the way the average collapse fraction $f(>m_{\rm min}|\delo,M_0)$ and the subsequent first crossing distribution are computed. 

One discrepancy when invoking the excursion set framework above to identify bubbles is that the calculation begins with the \emph{average} collapse fraction $f(>m_{\rm min}|\delo,M_0)$, so that all the stochasticity in the problem derives from the sequence of random variables $\delo(M_0)$. Clearly, however, the collapse fraction $\sum_hm_h/M_0$ in a given region of \emph{fixed} $(\delo,M_0)$ is itself a stochastic quantity, since, e.g., the number of sources in a given region is stochastic. This additional stochasticity, which is ignored in most excursion set models of \textsc{Hii} bubbles, would not be particularly problematic if it were uncorrelated with the stochasticity in $\delo$. E.g., \citet{fmh06} discuss how one can account for a stochastic collapse fraction deriving from Poisson distributed source number counts. In reality, however, the stochasticity of source number counts is neither Poisson nor uncorrelated with the large scale overdensity \citep{sl99a,sl99b}. As we discuss below, this can have dramatic consequences for the bubble distribution predicted in excursion set models.

\subsection{Photon number conservation: A diagnostic of the validity of bubble models}
\label{sec:critique:subsec:diagnostic}
\noindent
A straightforward diagnostic of the relative importance of the correlation between source number counts and large scale overdensity is the extent to which mass (equivalently, ionizing photon number) is not conserved in any model of \textsc{Hii} bubbles. To start with, we provide a simple proof that \emph{any} model with perfect knowledge of the distribution of individual sources across bubbles \emph{must} conserve mass. We will do this by calculating the total ionized mass fraction in the universe in two different ways and obtaining the same answer. Later, we show that excursion set models that ignore the stochasticity of number counts \emph{do not} conserve mass in general. 

\begin{figure}
\centering
\includegraphics[width=0.5\textwidth]{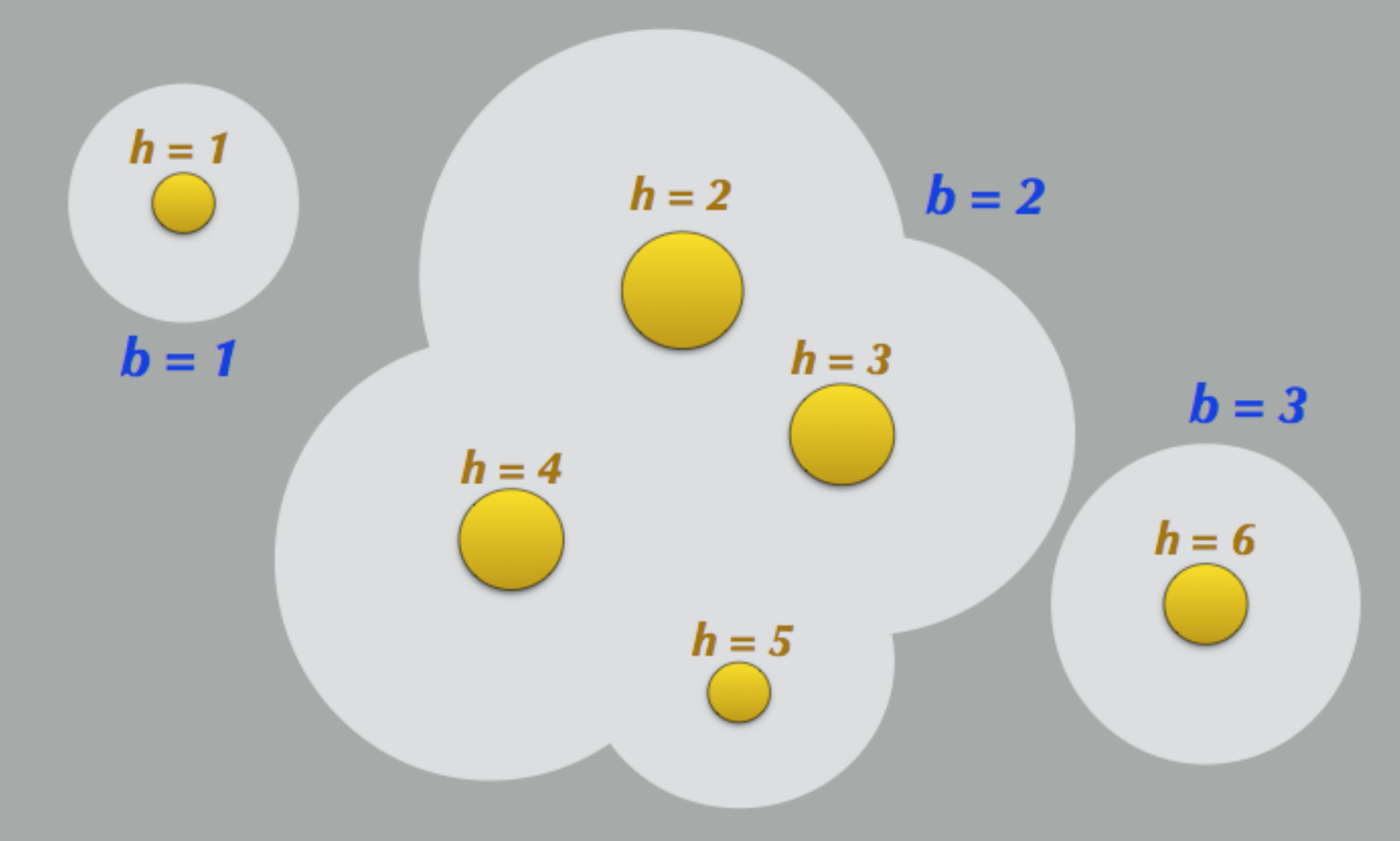}
\caption{Illustration of a partition of sources (yellow circles) into ionized bubbles (light shaded regions). The six sources labelled by $h$ are distributed among three bubbles labelled by $b$, with $b=1$ and $b=3$ sourced by a single source each while $b=2$ contains four sources. In this case we have $b_1=1$, $b_2=b_3=b_4=b_5=2$ and $b_6=3$, where $b_h$ is the label of the unique bubble sourced by the source $h$.}
\label{fig:bubbles-cartoon}
\end{figure}

Consider a box containing total mass $M_{\rm box}$. The box contains $N_{\rm src}$ sources with masses $m_h$, $h=1,2,..,N_{\rm src}$ which are partitioned among $N_{\rm bub}$ bubbles with ionized masses $M_b$, $b=1,2,..,N_{\rm bub}$. The source mass fraction in the box is $f_{\rm src} = \sum_{h=1}^{N_{\rm src}}m_h/M_{\rm box}$. Since each source $m_h$ ionizes a mass $\zeta m_h$, the \emph{ionized} mass fraction in the box is 
\be
f_{\rm ion} = \sum_{h=1}^{N_{\rm src}}\frac{\zeta m_h}{M_{\rm box}} = \zeta f_{\rm src}\,.
\label{f-ion-1}
\ee
Our basic assumption is that the sources are partitioned into bubbles, so that each source (labelled by $h$) sources a unique bubble (labelled by $b=b_h$), while a given bubble $b$ can be sourced by multiple sources $\{h_b\}$. Figure~\ref{fig:bubbles-cartoon} illustrates the situation. We can therefore rewrite the ionized fraction as follows:
\begin{align}
f_{\rm ion} 
&= \sum_{h=1}^{N_{\rm src}}\frac{\zeta m_h}{M_{b_h}}\frac{M_{b_h}}{M_{\rm box}}\,\notag\\
&= \sum_{h=1}^{N_{\rm src}}\,\sum_{b=1}^{N_{\rm bub}}\frac{\zeta m_h}{M_b}\,\delta_{b,b_h}\,\frac{M_b}{M_{\rm box}}\,\notag\\
&= \sum_{b=1}^{N_{\rm bub}}\frac{M_b}{M_{\rm box}}\,\left\{\sum_{h=1}^{N_{\rm src}}\frac{\zeta m_h}{M_b}\,\delta_{b,b_h}\right\}\notag\\
&= \sum_{b=1}^{N_{\rm bub}}\frac{M_b}{M_{\rm box}}\,\left\{\sum_{h_b}\frac{\zeta m_{h_b}}{M_b}\right\}\notag\\
&= \sum_{b=1}^{N_{\rm bub}}\frac{M_b}{M_{\rm box}}\notag\\
&= Q_{\rm Lag}\,.
\label{f-ion-2}
\end{align}
In the first line we simply multiplied and divided by the ionized mass of the unique bubble sourced by each source $h$. In the second line we introduced a sum over all bubbles, with a Kronecker delta to select the appropriate bubble for each source. In the third line we exchanged the order of the two summations, and in the fourth line we used the Kronecker delta to convert the sum over all sources into a sum over the sources of the bubble $b$. The quantity in braces is unity because of mass (or photon number) conservation: the numerator summed over $h_b$ gives the total ionized mass inside the bubble, which is equal to the bubble mass $M_b$ by definition. The last equality gives the definition of the quantity $Q_{\rm Lag}$, the Lagrangian ionized mass fraction, as a summation (or integral) over all bubbles of the ionized mass fraction in bubbles; this is a standard output of excursion set models of bubble mass distributions. 

This demonstrates the (rather obvious) result that in a model where each source of mass $m$ ionizes a mass $\zeta m$, \emph{and} we have perfect knowledge of the partition of sources into bubbles, we must have $\zeta f_{\rm src} = Q_{\rm Lag}$. The reason for the detailed calculation was to highlight the role of the knowledge of the partition, which appears as the list $\{b_h\}_{h=1}^{N_{\rm src}}$ giving the unique bubbles sourced by each source $h$. (Notice that this list will have repetitions.)

Now consider a traditional excursion set model (e.g., as proposed by FZH04) which uses the average source mass fraction in regions of fixed mass $M_0$ and overdensity $\delo$ to define an \textsc{Hii} bubble as the largest region satisfying $\zeta f(>m_{\rm min}|\delo,M_0) \geq 1$. In detail, this model tracks the fluctuations in $f(>m_{\rm min}|\delo,M_0)$ (due to fluctuations in $\delo$) as $M_0$ is decreased from large values. A key shortcoming of this approach is that the \emph{number} of sources in any given region is implicitly set equal to the average number at fixed $\delo$ and $M_0$: $N_{\rm src} = \avg{N_{\rm src}|\delo,M_0} = \int^{M_0}_{m_{\rm min}}\der m\,(M_0/m)\,f(m|\delo,M_0)$. This number can fluctuate up and down depending on the behaviour of $\delo$, and is therefore \emph{not guaranteed} to be monotonically decreasing as $M_0$ decreases. (That $N_{\rm src}$ \emph{must} be non-increasing as $M_0$ decreases is obvious if one considers sitting at the center of the box in the previous calculation and putting down concentric spheres of decreasing radius.) In other words, the fluctuations of source \emph{number} counts in this model are allowed to violate number (and hence mass and photon number) conservation. It is easy to see that this is a direct consequence of ignoring the partitioning information which we used in the calculation above.

This shows that traditional excursion set models, when used to calculate bubble distributions, will violate mass/photon number conservation in general. This shortcoming of the excursion set approach has been known for a considerably long time \citep{sl99a,sl99b}; the fact that excursion set models have nevertheless been very successful in modelling halo abundances and clustering is largely because this violation of mass conservation does not show up when considering average quantities, but only when dealing with (source number) fluctuations. As such, the construction of halo merger histories can be quite sensitive to this issue; Sheth \& Lemson addressed precisely this problem in their papers and showed how one might modify the excursion set framework to account for mass/number conservation when modelling merger trees. As we have seen, the problem of defining \textsc{Hii} bubbles is another prime example requiring knowledge of fluctuating numbers of objects, and we will show below that the techniques employed by Sheth \& Lemson are also useful in this setting.

Before we do so, it is worth exploring the FZH04 model to determine the extent to which it violates photon number conservation. We emphasize, however, that the qualitative nature of these results is generic and also applies to all other excursion set based bubble distributions to date.

\subsection{$\zeta f_{\rm src}$ Versus $Q_{\rm Lag}$ in the FZH04 model}
\label{sec:critique:subsec:phNCinFZH}
\noindent
The FZH04 model builds on the sharp-$k$ filtered excursion set model of \citet{bcek91} and starts by setting the average conditional source mass fraction to 
\be
f(>m_{\rm min}|\delo,M_0) = \erfc{\frac{\delc(z)-\delo}{\sqrt{2(s_{\rm min}-S_0)}}}\,,
\label{fsrc-FZH}
\ee
where $\delc(z)$ is the critical linear density for spherical collapse at redshift $z$, linearly extrapolated to present day, and we defined $s_{\rm min} = \sig^2(m_{\rm min})$ and $S_0 = \sig^2(M_0)$, with $\sig^2(m) = \avg{\delta(m)^2}$ being the variance of the initial matter density fluctuations smoothed on scale $r\propto m^{1/3}$ and linearly extrapolated to present day. The ionized fraction of the universe follows from setting $\delo,S_0\to0$ and multiplying by $\zeta$,
\be
\zeta f_{\rm src} = \zeta\,\erfc{\delc(z)/\sqrt{2s_{\rm min}}}\,.
\label{fion-FZH}
\ee
Since the conditional mass fraction \eqref{fsrc-FZH} is a monotonic function of \delo, the condition $\zeta f(>m_{\rm min}|\delo,M_0) \geq 1$ is equivalent to  $\delo\geq B_{\textsc{Hii}}(M_0;z)$ where
\be
B_{\textsc{Hii}}(M_0;z) \equiv \delc(z) - \sqrt{2}\, K(\zeta)\sqrt{s_{\rm min} - S_0}\,,
\label{BHII-FZH}
\ee
with $K(\zeta)=\erfcinv{1/\zeta}$. The problem is now identical to the first-crossing problem for random walks in the smoothed matter density that \citet{bcek91} solved for the halo mass function, except that the ``barrier'' which must be crossed is given by the function $B_{\textsc{Hii}}(M_0;z)$ rather than $\delc(z)$. The resulting first-crossing distribution $f_{\textsc{Hii}}(M_0)\der M_0$ then gives the ionized mass fraction in bubbles of mass $M_0$, and integrating this gives us $Q_{\rm Lag}$,
\be
Q_{\rm Lag} = \int_{\zeta m_{\rm min}}^\infty\der M_0\,f_{\textsc{Hii}}(M_0)\,,
\label{QLag-def}
\ee
where the lower limit of the integral follows from the requirement that the smallest source mass is $m_{\rm min}$, so that the smallest bubble can have mass $\zeta m_{\rm min}$.

\begin{figure}
\centering
\includegraphics[width=0.45\textwidth]{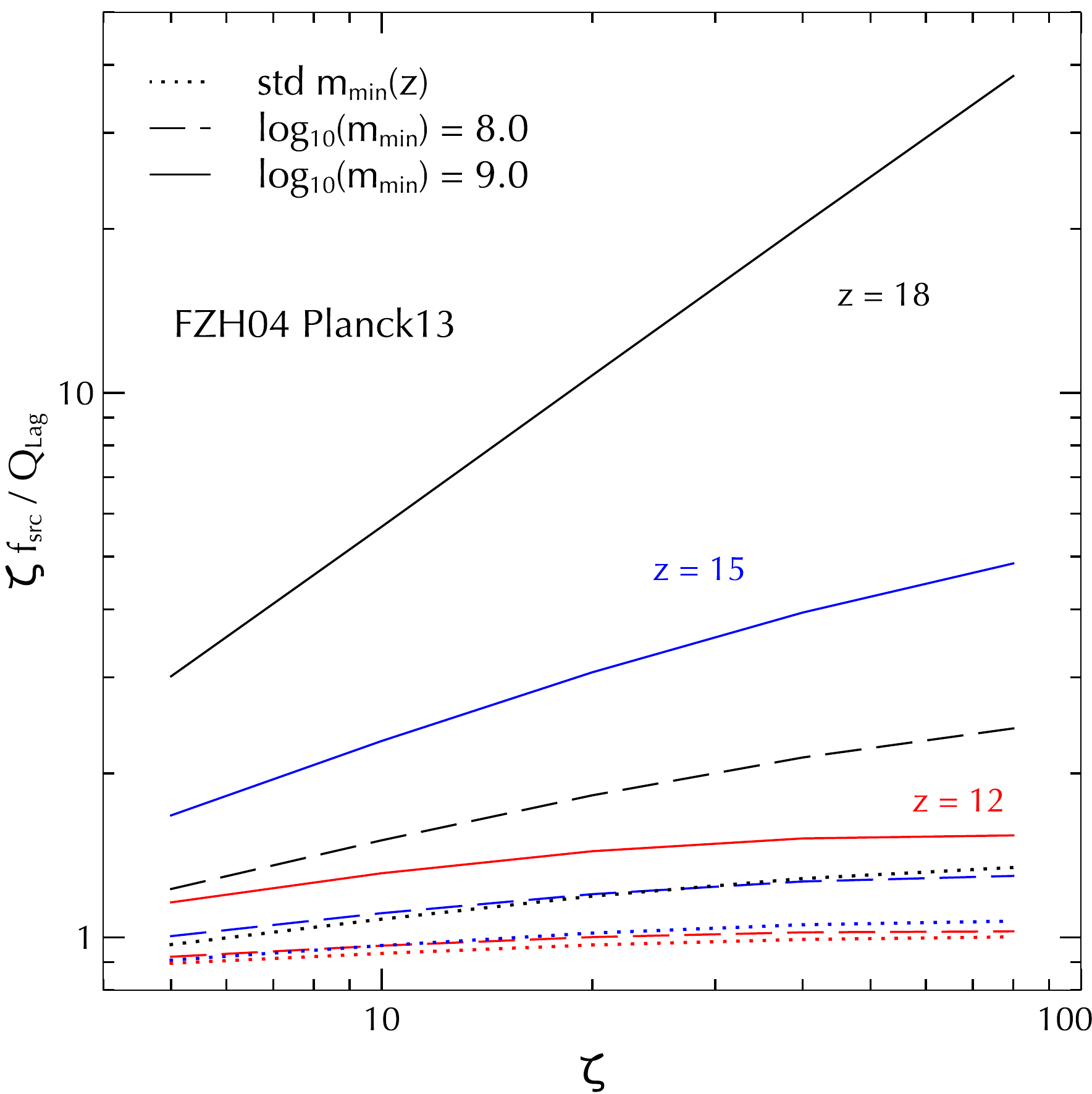}
\caption{Ratio of $\zeta f_{\rm src}$ (equation~\ref{fion-FZH}) and $Q_{\rm Lag}$ in the linear barrier approximation of FZH04 \citep[equation 12 of][]{pc14} for multiple redshifts and choices of $m_{\rm min}$, as a function of $\zeta$. The curves labelled ``std'' use the standard choice where $m_{\rm min}$ is the mass associated with virial temperature $10^4$K.}
\label{fig:QLagVszetafc-FZHlin}
\end{figure}

Although the use of a filter that is sharp in Fourier space considerably simplifies the problem (because one now deals with random walks with uncorrelated steps), there is unfortunately no analytical solution for the first crossing of the barrier \eqref{BHII-FZH}. FZH04 therefore advocated using an approximation to the barrier which is linear in $S_0$,
\begin{align}
B_{\textsc{Hii}{\rm ,lin}}(M_0;z) &\equiv B_0 + B_1\,S_0\,,\label{BHII-FZHlin}\\ 
B_0 &= \delc(z) - K(\zeta)\sqrt{2s_{\rm min}}\,,\notag\\
B_1 &= K(\zeta)/\sqrt{2s_{\rm min}}\,,
\end{align}
for which there \emph{is} an analytical solution due to \citet{sheth98}. As pointed out by \citet[][see their equation 12]{pc14}, the resulting expression for $Q_{\rm Lag}$ is different than \eqn{fion-FZH}. Figure~\ref{fig:QLagVszetafc-FZHlin} shows the ratio $\zeta f_{\rm src}/Q_{\rm Lag}$ for this linear barrier approximation as a function of $\zeta$ for several redshifts, and we see that the ratio can be substantially larger than unity at high redshifts for large $\zeta$.

\begin{figure}
\centering
\includegraphics[width=0.45\textwidth]{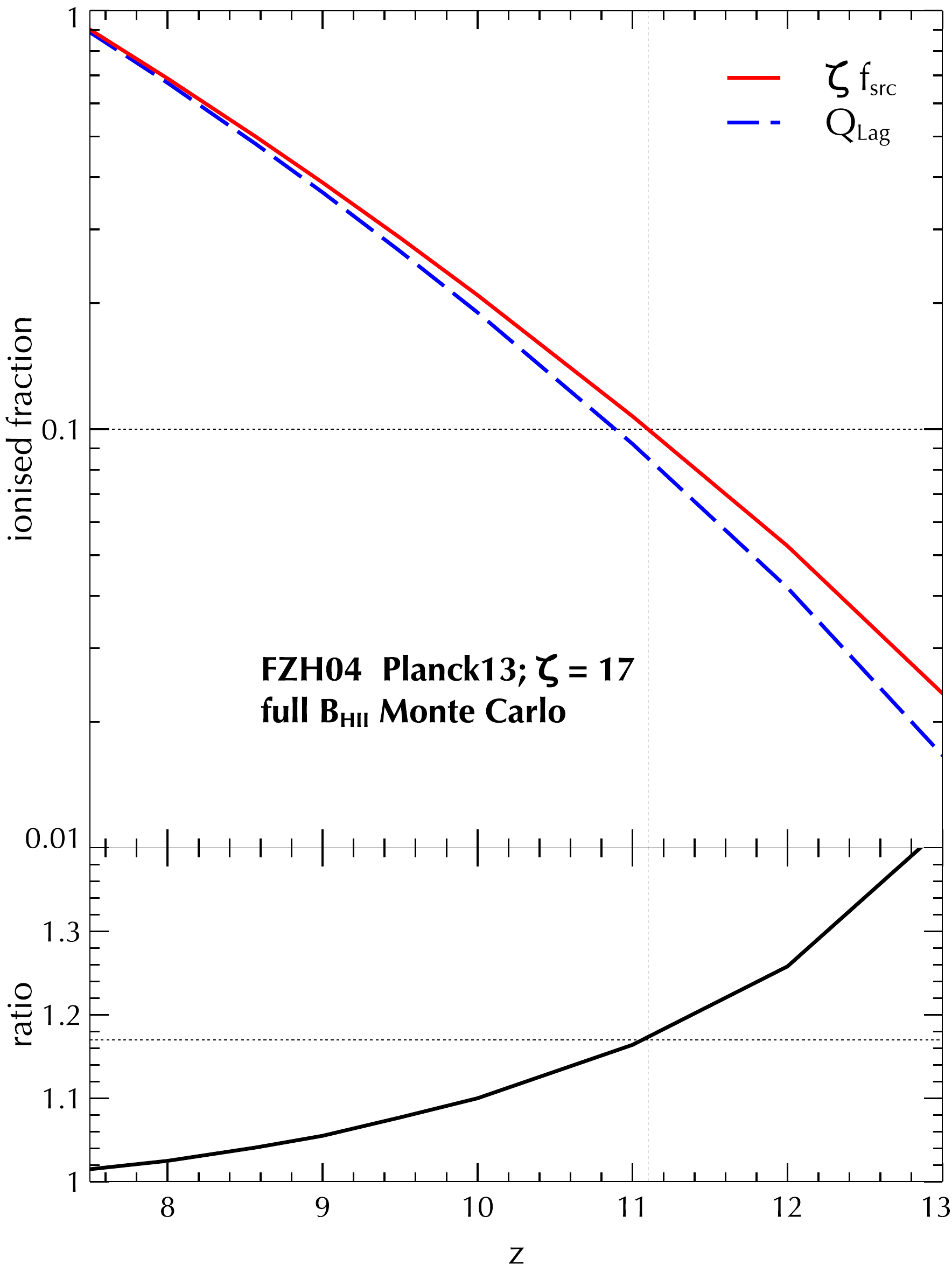}
\caption{$\zeta f_{\rm src}$ (equation~\ref{fion-FZH}) and $Q_{\rm Lag}$ as a function of redshift using the full Monte Carlo solution for the first crossing of the barrier \eqref{BHII-FZH}, calculated using $10^6$ sharp-$k$ walks with Planck13 initial conditions and setting $m_{\rm min}=10^8\Mh$. For these settings, our choice of constant $\zeta=17$ leads to an electron optical depth $\tau_{\rm el}=0.066$, consistent with the recent constraint from \citet{Planck13-XVI-cosmoparam}. The \emph{top panel} shows the $\zeta f_{\rm src}$ and $Q_{\rm Lag}$ separately, while the \emph{bottom panel} shows the ratio $\zeta f_{\rm src}/Q_{\rm Lag}$. The vertical dotted line marks the redshift where $\zeta f_{\rm src}=0.1$; we see in the bottom panel that the discrepancy between $\zeta f_{\rm src}$ and $Q_{\rm Lag}$ is $\gtrsim15$ per cent even at this late stage.}
\label{fig:QLagVszetafc-FZHhistory}
\end{figure}

One might argue \citep{fzh04,zahn+07} that this is either a consequence of the linear barrier approximation or simply due to using unrealistic values of $\zeta$ and redshift, and that the situation might improve if we had the exact solution to the FZH04 first crossing problem using a realistic reionization history. Our previous arguments suggest otherwise, and Figure~\ref{fig:QLagVszetafc-FZHhistory} explicitly shows that $\zeta f_{\rm src}\neq Q_{\rm Lag}$ even when $Q_{\rm Lag}$ is calculated using a full numerical solution of the first crossing problem of the barrier \eqref{BHII-FZH} by sharp-$k$ filtered walks\footnote{See \citet{bcek91} for a description of the algorithm used to determine the first crossing distribution numerically.}. The Figure shows $\zeta f_{\rm src}$ and $Q_{\rm Lag}$ as a function of redshift, as predicted by the Monte Carlo solution of the FZH04 model when using cold dark matter (CDM) ICs with a Planck13 cosmology to determine the $\sig^2(m)$ relation. We set $m_{\rm min}=10^8\Mh$ and demand that the model yield a value of electron scattering optical depth $\tau_{\rm el}=0.066$, consistent with recent Planck constraints \citep{Planck13-XVI-cosmoparam}; this gives us the value $\zeta=17$ used in the Figure. The vertical dotted line indicates the redshift at which $\zeta f_{\rm src}=0.1$, and the bottom panel shows that $\zeta f_{\rm src}$ and $Q_{\rm Lag}$ are discrepant at $\gtrsim15$ per cent even at this late stage in a realistic setting.

\begin{figure*}
\centering
\includegraphics[width=0.45\textwidth]{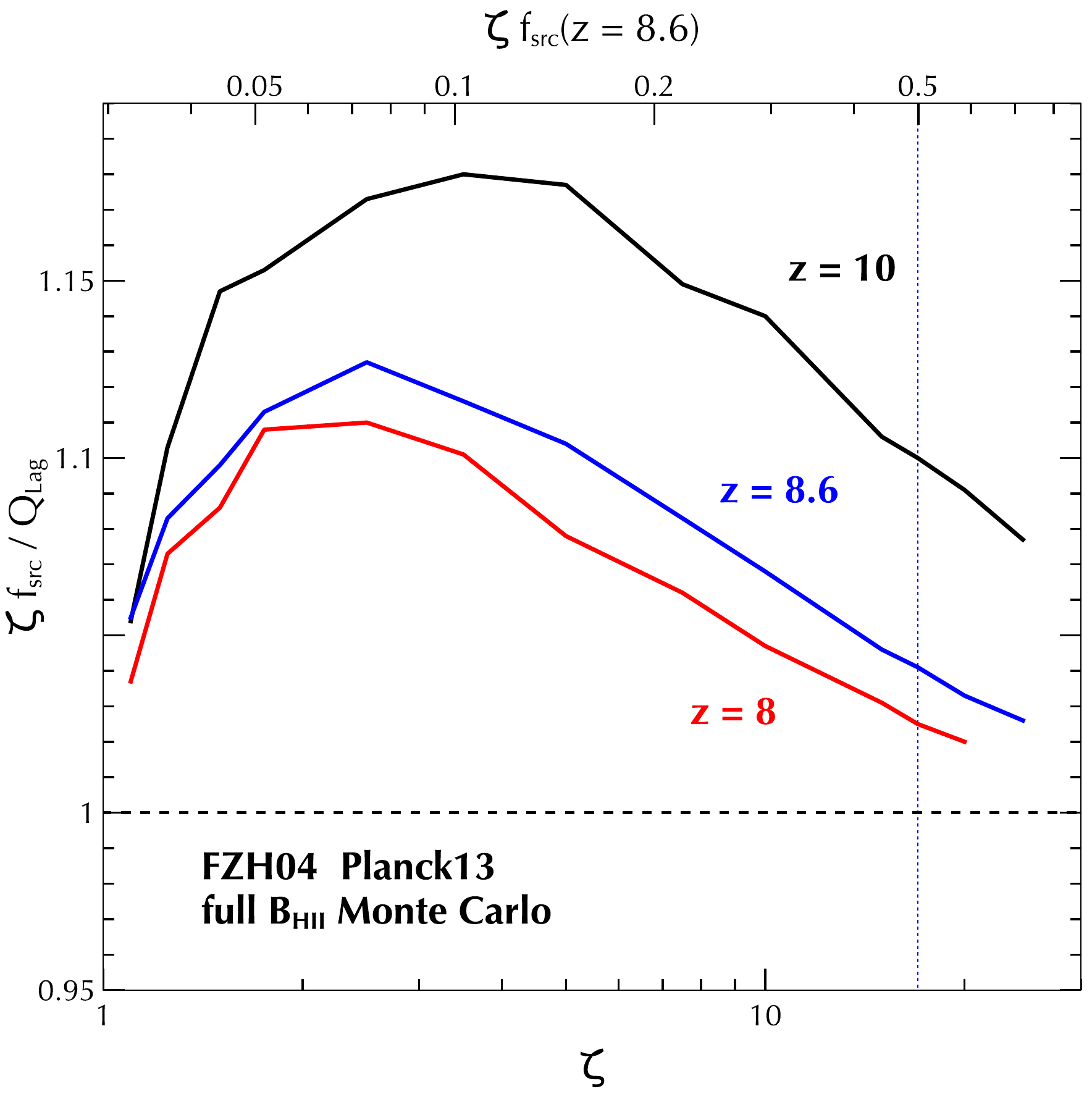}
\includegraphics[width=0.45\textwidth]{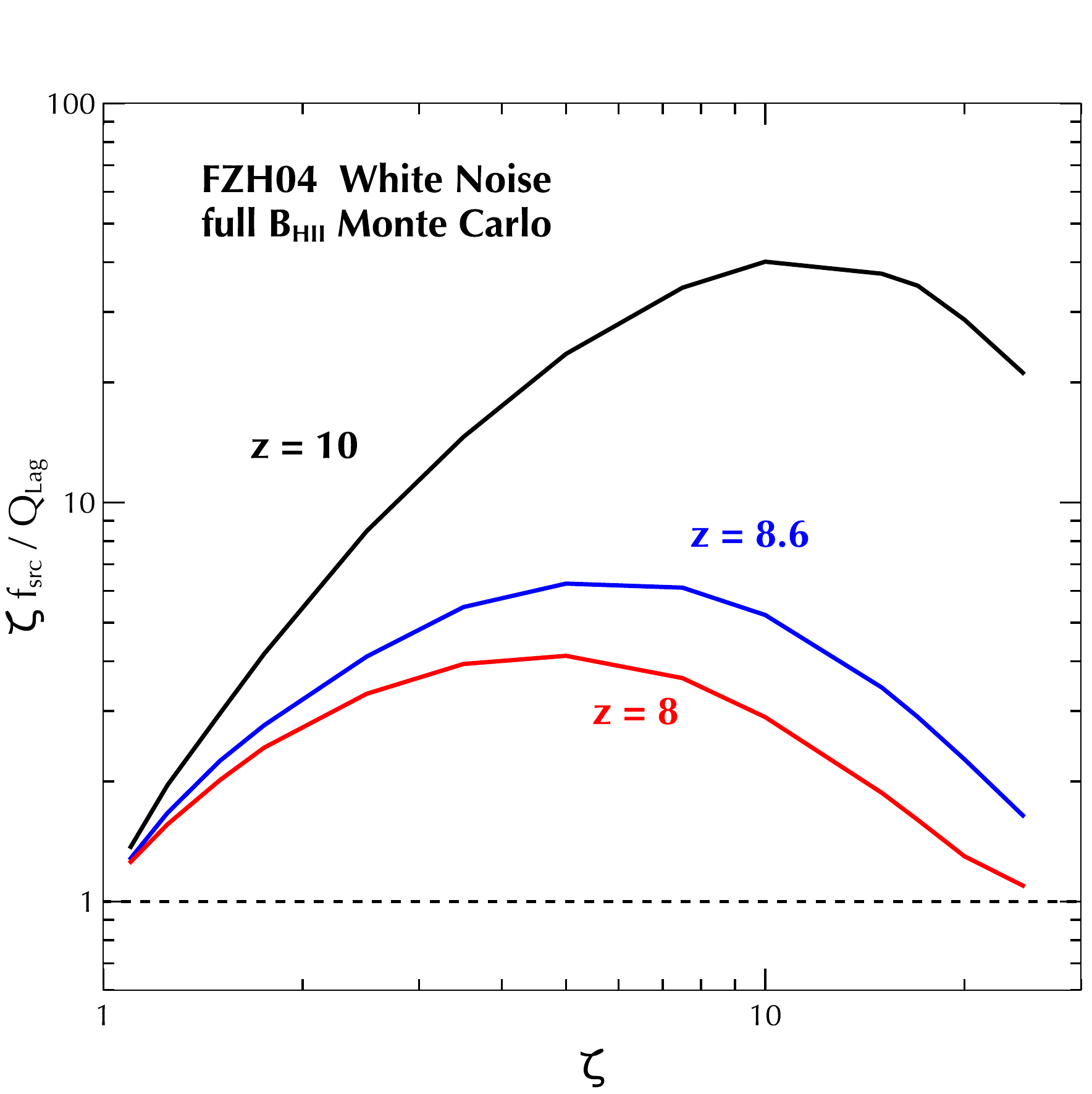}
\caption{Ratio of $\zeta f_{\rm src}$ (equation~\ref{fion-FZH}) and $Q_{\rm Lag}$ using the full Monte Carlo solution for the first crossing of the barrier \eqref{BHII-FZH}, calculated using $10^6$ sharp-$k$ walks for multiple redshifts as a function of $\zeta$, setting $m_{\rm min}=10^8\Mh$. \emph{(Left panel):} $\Lambda$CDM initial conditions assuming a Planck13 cosmology. The top axis indicates the value of $\zeta f_{\rm src}$ at $z=8.6$, at which time the `realistic' model in Figure~\ref{fig:QLagVszetafc-FZHhistory} reaches $\zeta f_{\rm src}=0.5$ (shown as the vertical blue dotted line). \emph{(Right panel):} Initial conditions assume a white noise power spectrum normalised so as to match the Planck13 value of $s_{\rm min}$ for $m_{\rm min}=10^8\Mh$, leading to identical ionization histories $\zeta f_{\rm src}(z)$ at the same value of $\zeta$ across the two panels. Notice the difference in scale on the vertical axis between the two panels.}
\label{fig:QLagVszetafc-FZHfull}
\end{figure*}

Purely from the model-building perspective, it is then useful to study this behaviour as a function of the model parameters, regardless of how realistic their values might be. Similar to Figure~\ref{fig:QLagVszetafc-FZHlin}, Figure~\ref{fig:QLagVszetafc-FZHfull} explores the behaviour of the full numerical solution as a function of $\zeta$.
The left panel shows the ratio $\zeta f_{\rm src}/Q_{\rm Lag}$ for Planck13 $\Lambda$CDM ICs, for three values of redshift chosen such that the central value $z=8.6$ corresponds to $50\%$ ionization for the realistic value $\zeta=17$ (see Figure~\ref{fig:QLagVszetafc-FZHhistory}). The top axis shows the value of $\zeta f_{\rm src}$ at $z=8.6$ for the other values of $\zeta$. We have checked that all parameter combinations shown in our plots correspond to $\zeta f_{\rm src} < 1$.

The right panel shows the same ratio using white noise ICs ($P(k)=\,$constant). We normalize the white noise power spectrum so as to match the CDM value of $s_{\rm min}$ at $m_{\rm min}=10^8\Mh$. Since $f_{\rm src}$ depends on the power spectrum only though the value of $s_{\rm min}$ (see equation~\ref{fion-FZH}), the ionization history $\zeta f_{\rm src}(z)$ is identical for corresponding values of $\zeta$ across the two panels. We see that white noise ICs lead to a severe disagreement between $\zeta f_{\rm src}$ and $Q_{\rm Lag}$; this will be important below.

We also see that we always have $Q_{\rm Lag} < \zeta f_{\rm src}$ for the full FZH04 solution. In fact, one can go further and prove the following curious result: if the barrier \eqref{BHII-FZH} is artificially (and unphysically) extended below $M_0=\zeta m_{\rm min}$ all the way to $M_0=m_{\rm min}$, then the resulting $Q_{\rm Lag}$ exactly equals $\zeta f_{\rm src}$ (see Appendix~\ref{app:zetafcoll}). Purely on algebraic grounds, then, we see that the physically correct $Q_{\rm Lag}$, which only gets contributions from $M_0 > \zeta m_{\rm min}$, \emph{must} be smaller than $\zeta f_{\rm src}$. There are two limits in which $Q_{\rm Lag}$ approaches $\zeta f_{\rm src}$, however. The first is when $\zeta\to1$, where $\zeta m_{\rm min}\to m_{\rm min}$ and the argument above applies. The second is at large $\zeta$ where both $\zeta f_{\rm src}$ and $Q_{\rm Lag}$ approach unity. Figure~\ref{fig:QLagVszetafc-FZHfull} shows how these limits are approached in various cases, with a maximum value for the ratio in between.

At this point, it is worth asking whether such a violation of photon number conservation exists also in the excursion set based semi-numerical simulations of ionized bubbles. It turns out that it does, though the reasons may not be same as those for the analytical approaches. \citep[As already discussed by][currently popular semi-numerical schemes are substantially different than analytical approaches in their implementation and assumptions.]{pc14} We discuss the semi-numerical simulations in Appendix \ref{app:seminum}, keeping the main focus of the paper solely on analytical models.

\section{A photon number conserving model of ionized bubbles}
\label{sec:numconsbub}
\noindent
To solve the problem of mass or photon number conservation, we need a model that explicitly counts individual sources in a region, not just their overall mass fraction. This would be possible if we could partition a given region of total mass $M_0$ and overdensity \delo\ into halos. \citet{sl99b} describe an algorithm that does exactly this, provided the ICs are generated from a white noise power spectrum. White noise ICs ensure that disjoint regions are statistically independent, with $\sigma^2\propto 1/V\propto 1/M$; this allows the construction of self-consistent, mass conserving partitions. This statistical independence is lost when using realistic CDM ICs; nevertheless, the same partition algorithm can be used at the cost of reduced accuracy in reproducing mean mass fractions. 

\begin{figure*}
\centering
\includegraphics[width=0.85\textwidth]{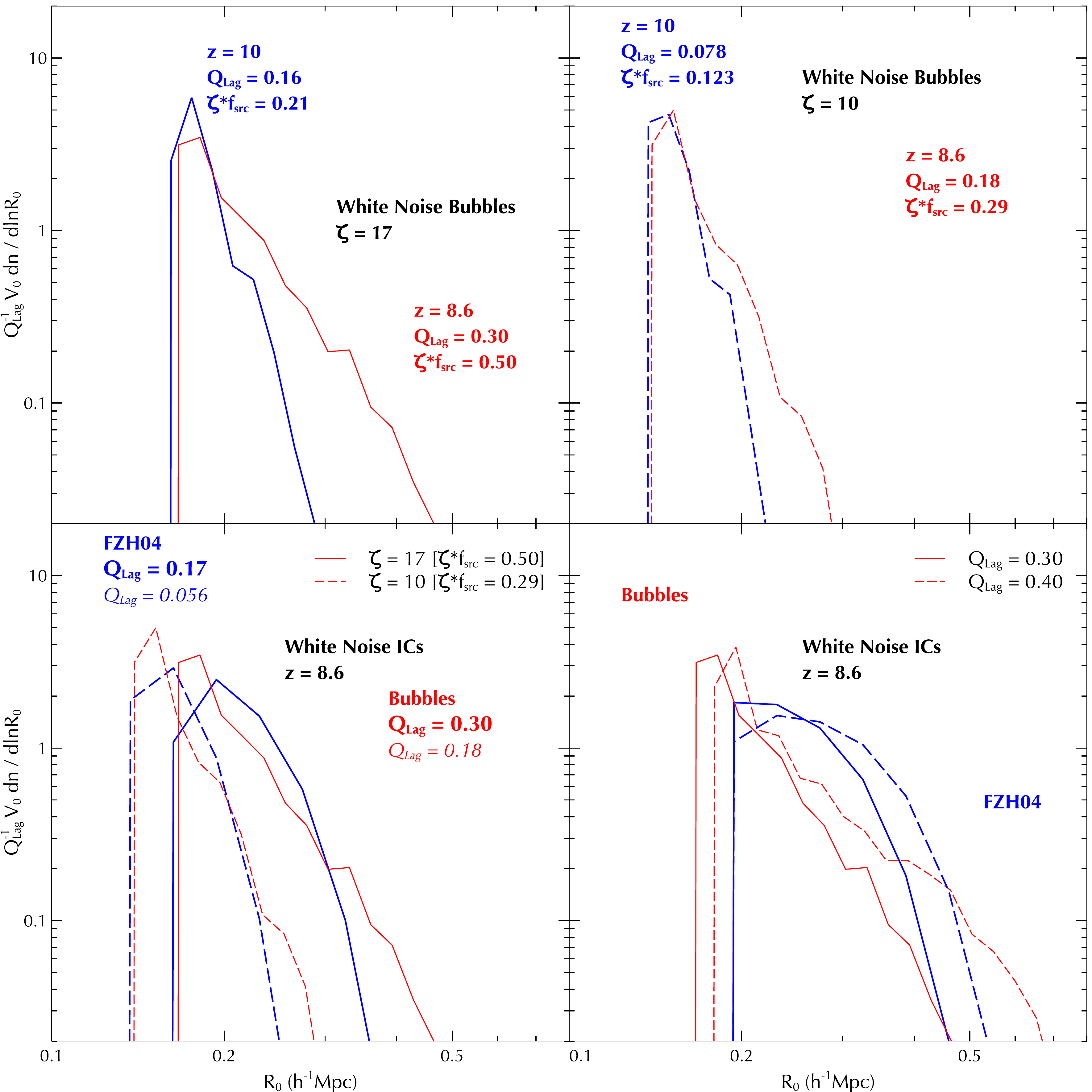}
\caption{Size distribution of bubbles using white noise ICs with $m_{\rm min}=10^8\Mh$. \emph{(Top row):} The predictions for our model at two redshifts $z=8.6$ (red; larger bubble sizes) and $z=10$ (blue; smaller bubble sizes) for two values of $\zeta$, $17$ (\emph{left panel}, solid) and $10$ (\emph{right panel}, dashed). The values of $\zeta f_{\rm src}$ and the corresponding values of $Q_{\rm Lag}$ are given in the labels. Note that $\zeta=17$ with $m_{\rm min}=10^8\Mh$ corresponds to a realistic ionization history $\zeta f_{\rm src}(z)$ (see Figure~\ref{fig:QLagVszetafc-FZHhistory}). \emph{(Bottom left panel):} Comparison of this work with FZH04 at $z=8.6$ for the same ICs and values of $\zeta$. The red curves (with larger bubble sizes) are repeated from the top row while the blue solid and dashed curves (somewhat smaller bubble sizes) give the FZH04 results for $\zeta=17$ and $\zeta=10$, respectively. The $Q_{\rm Lag}$ values in each model are given in the labels, in boldface for $\zeta=17$ and in italics for $\zeta=10$. Our model predicts a $Q_{\rm Lag}$ which is within $\sim 40\%$ of $\zeta f_{\rm src}$, while the $Q_{\rm Lag}$ for FZH04 is a factor $3$-$5$ smaller. \emph{(Bottom right panel):} Comparison of this work (red) with FZH04 (blue) at $z=8.6$ for fixed values of $Q_{\rm Lag}$. The values of $\zeta$ for both models have been adjusted to obtain the same $Q_{\rm Lag}$, with the results shown as the solid (smaller $Q_{\rm Lag}$) and dashed curves (larger $Q_{\rm Lag}$). 
In all cases, the results of our algorithm are based on $20,000$ walks while the FZH04 results are based on $10^6$ walks.}
\label{fig:sizedist}
\end{figure*}

We describe the Sheth-Lemson (henceforth SL99) partition algorithm in Appendix~\ref{app:SL99}. In this section, we assume that this algorithm is at hand and allows us to generate a list of source masses $\{m_h\}$, $h=1,2,\ldots,N_{\rm src}$ for any region of mass $M_0$ and overdensity \delo, where the individual masses $m_h$ as well as the total number of sources $N_{\rm src}$ are generated stochastically. We now show how this algorithm can be used as a workhorse in a Monte Carlo approach to count ionized bubbles while conserving photon number.

Our basic strategy is to first populate a very large region (a proxy for ``the whole universe'') with sources, starting from a small sphere at the center (which we take to enclose a mass $\zeta m_{\rm min}$, the smallest expected bubble size) and proceeding outwards in concentric shells. Since the innermost sphere and the concentric shells form a set of disjoint regions, the SL99 algorithm can be used independently in each of them. The resulting sources are tagged not only by their masses but also by the shell that they occupy. The algorithm requires the total mass and overdensity of each shell as an input. We generate these self-consistently in two steps: 
\begin{enumerate}
\item We first generate a random walk in spherical overdensity $\{\delta_j,V_j\}$ where $V_j$ is the volume of the $j^{\rm th}$ concentric \emph{sphere} (not shell) counting from inside out, and $\delta_j$ is its overdensity in the ICs. This can be done using standard techniques for a chosen filter and power spectrum \citep{bcek91,mp12}. We use a spherical TopHat filter and either white noise or power law ICs.
\item Knowing the volume and overdensity of this series of spheres allows us to calculate the overdensity and mass of the corresponding shells $\{\Delta\delta_j,\Delta M_j\}$: we have $\Delta M_j = M_j-M_{j-1}$ and $\Delta\delta_j = (M_j\delta_{j}-M_{j-1}\delta_{j-1})/\Delta M_j$ at leading order in overdensity, with $\Delta M_1 = M_1$ and $\Delta\delta_1 = \delta_1$ corresponding to the innermost sphere which we treat as the first ``shell''.
\end{enumerate}
The sequence $\{\Delta\delta_j,\Delta M_j\}$ is fed to the SL99 algorithm which returns a list of source masses $\{m_{h,j}\}$ in each shell $j$, with $h=1,2,\ldots,N_{\rm src,j}$. We now proceed \emph{outside in} and calculate the ionized mass fraction $\sum_h\zeta m_h/M_j$ in each \emph{sphere}, where the sum runs over all sources enclosed by the $j^{\rm th}$ sphere (not shell). The size of the largest sphere for which the ionized fraction exceeds unity is recorded as a bubble size. If no sphere satisfies this condition, this walk is declared as having ``not crossed'' the ionization barrier. 

We repeat this procedure many times, building up the distribution of bubble sizes. It should be clear from our earlier discussion that the individual bubbles explicitly conserve photon number. The total ionized fraction $Q_{\rm Lag}$ is given by the fraction of simulated walks that crossed the barrier. We compare the outcome of this algorithm with the FZH04 result in Figure~\ref{fig:sizedist}. The top row of the Figure shows the bubble size distribution of our algorithm for two different redshifts (colour-coded) and two different values of $\zeta$ (the left and right panels), for the same white noise ICs used in the right panel of  Figure~\ref{fig:QLagVszetafc-FZHfull}. One can see that the model predicts $Q_{\rm Lag} < \zeta f_{\rm src}$, however, the difference between the two quantities is within $\lesssim 40\%$. This should be compared with the dramatically large discrepancies seen in the right panel of  Figure~\ref{fig:QLagVszetafc-FZHfull}. The fact that we still predict $Q_{\rm Lag} < \zeta f_{\rm src}$ with white noise ICs (for which we might have expected perfect agreement) is almost certainly due to our adoption of the excursion set ansatz which implicitly equates the results of a large number of simulations tracking a single bubble each (as we do when predicting $f_{\rm ion}$ as $Q_{\rm Lag}$) with counting many bubbles in a single box (as assumed when setting $f_{\rm ion}=\zeta f_{\rm src}$). Fixing this will require modifications to our algorithm which are beyond the scope of this paper.

The bottom left panel of Figure~\ref{fig:sizedist} compares the bubble size distribution at a single redshift with the FZH04 calculation using the same white noise ICs. This shows that the improvement in the ratio $\zeta f_{\rm src}/Q_{\rm Lag}$ in our model as compared to FZH04 is achieved, in part, by generating a larger number of large bubbles. Note that the difference between the two models can be quite significant for relatively larger values of the bubble size.

An alternate way of comparing our approach to that of FZH04 would be to normalize the results to a fixed value of $Q_{\rm Lag}$ rather than $\zeta$. We do this by adjusting the value of $\zeta$ until the two models give the same $Q_{\rm Lag}$. The results are shown in the bottom right panel of Figure~\ref{fig:sizedist}. 
There are still considerable differences between the two models\footnote{We have checked, however, that the relative differences between the two models are rather sensitive to the choice of parameter values. E.g., for an `early reionization' scenario with $Q_{\rm Lag}(z=11)\simeq0.5$, we find that there is remarkably little difference in the shapes of the two bubble distributions for large bubbles, despite the fact that the two models require substantially different values of $\zeta$ to achieve this value of $Q_{\rm Lag}$. We leave a fuller exploration of the parameter space to future work.}.
Since the two models now have different $\zeta$, the size of the smallest bubble is also different; our model requires a smaller $\zeta$ to attain the same $Q_{\rm Lag}$ as FZH04, and hence tends to produce a larger number of small bubbles.

\begin{figure}
\centering
\includegraphics[width=0.45\textwidth]{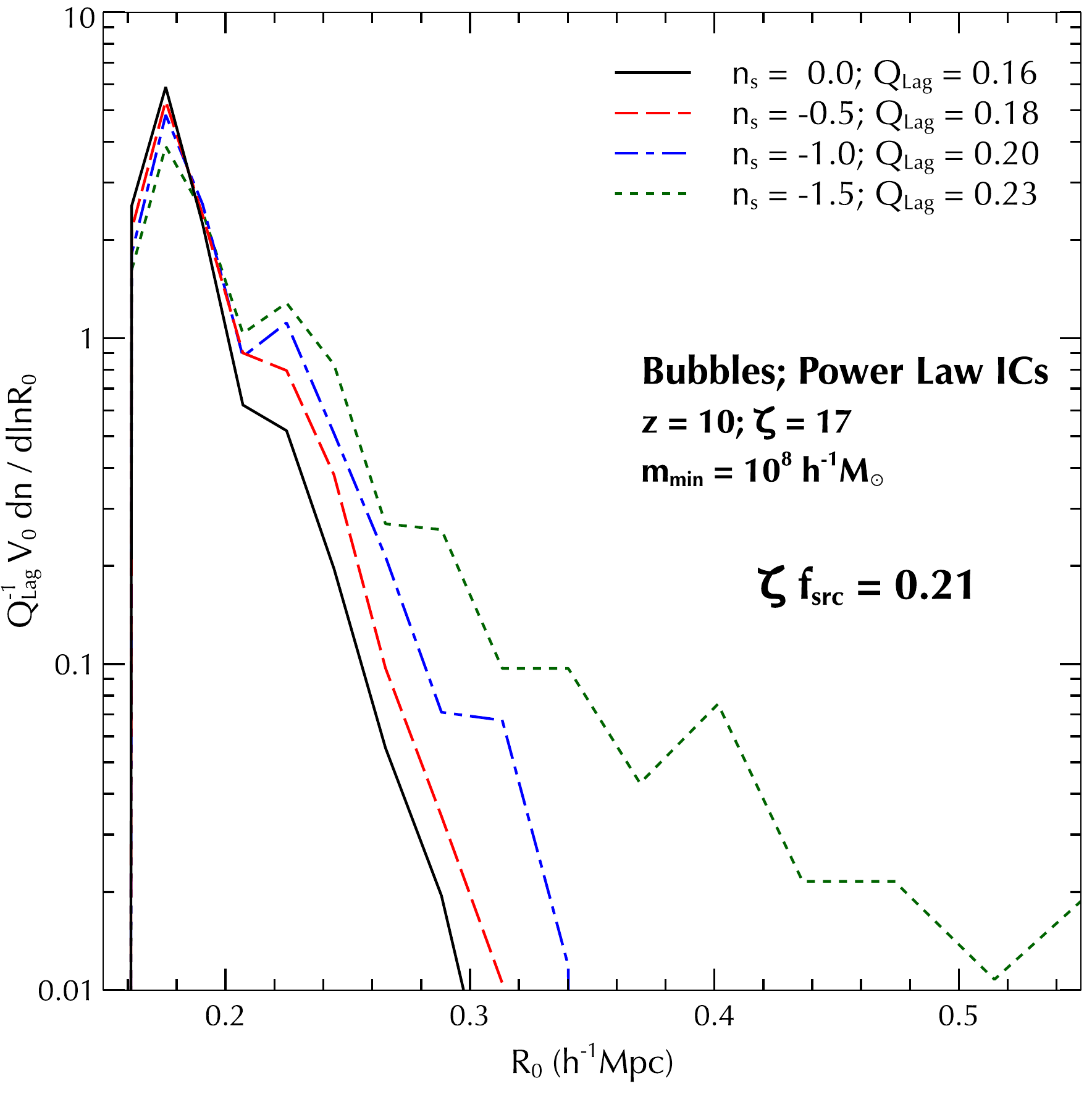}
\caption{Size distribution of bubbles in our algorithm for scale free power spectra $P(k) \propto k^{n_s}$, with different values of $n_s$, including $n_s=0$ which is white noise. The power spectra are all normalised so as to give the same value of $s_{\rm min}$ at $m_{\rm min}=10^8\Mh$, which means that they all lead to the same value of $\zeta f_{\rm src}$ for fixed $\zeta$. 
In particular, our choice of $\zeta$ is the one that leads to a realistic value of $\tau_{\rm el}$ as discussed in the text.}
\label{fig:sizedist-ns}
\end{figure}

As mentioned earlier, it is not straightforward to extend the partitioning algorithm of \citet{sl99b} to ICs other than white noise because the disjoint regions are no longer statistically independent. However, following one of the approaches advocated by \citet{sl99b}, we simply assume that the algorithm can be used directly for other ICs too. The results in this case will differ from the white noise IC results because the mass variance $\sigma^2(m)$ depends on the power spectrum. The disadvantage of this method is that it is no longer certain that the mean values generated by the algorithm will match the excursion set values; however, as was shown by \citet{sl99b}, the disagreement is not significant at least for scale free power spectra $P(k) \propto k^{n_s}$. The bubble size distribution for different values of $n_s$ is shown\footnote{Since computational costs increase for steeper spectra, our results used progressively fewer walks for decreasing values of $n_s$; we have checked that at least the values of $Q_{\rm Lag}$ are well converged in each case. For the same reason, we do not perform the analysis with CDM ICs.} in Figure~\ref{fig:sizedist-ns}, for $z=10$ and $\zeta=17$. The values of $Q_{\rm Lag}$ for different $n_s$ are quoted in the same Figure. 
As before, the power spectra are normalized so that the variance $s_{\rm min}=\sigma^2(m_{\rm min})$ matches the corresponding Planck13 $\Lambda$CDM value in each case, and the value of $\zeta f_{\rm src}$ is consequently independent of $n_s$. In particular, our choice of $\zeta$ is the one that leads to a realistic value of $\tau_{\rm el}$ as discussed previously.

From the Figure, we can see that as we decrease the value of $n_s$ from zero (which corresponds to the white noise ICs), the number of relatively large bubbles slowly increases. This leads to an increase in the value of $Q_{\rm Lag}$, giving a better match with the value of $\zeta f_{\rm src}$. 
The ratio $\zeta f_{\rm src}/Q_{\rm Lag}$ for our model with $n_s = 0,-0.5,-1,-1.5$ is, respectively, $1.35,1.14,1.03,0.93$. 
Correspondingly, we have checked that the FZH04 model with the same settings gives a ratio $\zeta f_{\rm src}/Q_{\rm Lag} \approx 36,10,4,2$, respectively, showing that our algorithm leads to a substantial improvement in these cases as well.
Since $n_s=-1.5$ is expected to be close to CDM ICs over typical scales of interest, the trends above then indicate that our algorithm would also lead to similar improvements for CDM ICs. In particular, since the actual FZH04 ratio for CDM ICs inferred from the left panel of Figure~\ref{fig:QLagVszetafc-FZHfull} is much closer to unity ($\sim10$ per cent discrepant), our algorithm should also lead to only a minor discrepancy for CDM ICs.

\section{Conclusions}
\label{sec:conclude}
\noindent
In excursion set based models of ionized bubbles \citep[e.g.,][FZH04]{fzh04} it is expected that the ratio, $\zeta f_{\rm src}$, of the number of ionizing photons produced by all sources (compensated for recombinations in the medium) and the number of hydrogen atoms should be equal to $Q_{\rm Lag}$, the fraction of mass contained within all ionized bubbles. It is however seen that $Q_{\rm Lag} \neq \zeta f_{\rm src}$ in general \citep[c.f., e.g., FZH04 and][see also section~\ref{sec:critique:subsec:phNCinFZH} and Appendix~\ref{app:seminum}]{zahn+07}, so that photons are being spuriously lost or gained in these models (for the FZH04 model they are always lost). This non-conservation of photon number is worrisome because it indicates that the models are missing some key physical or statistical ingredient\footnote{We argued that previous `solutions' of this problem were based on erroneous calculations (Appendix~\ref{app:zetafcoll}).}.

In this paper we demonstrated that this non-conservation arises because excursion set models use only the average mass fraction in a region and do not account for the exact (stochastically fluctuating) number and masses of the sources (section~\ref{sec:critique:subsec:diagnostic}). This insight allowed us to build a Monte Carlo model of bubble growth that explicitly counts the number of sources in a region and is therefore expected to conserve photon number (section~\ref{sec:numconsbub}). Although this model \citep[based on the partitioning algorithm presented by][see Appendix~\ref{app:SL99}]{sl99b} is formally accurate only for white noise initial conditions (ICs), we showed that its \emph{ad hoc} extension to scale free power spectra also leads to reasonable results, with $Q_{\rm Lag}$ becoming progressively larger for steeper spectra at the same value of $\zeta f_{\rm src}$.
In all cases, the match between $Q_{\rm Lag}$ and $\zeta f_{\rm src}$ using our algorithm is substantially improved as compared to that using the FZH04 algorithm with the same ICs. In particular, at fixed $\zeta$ our model returns a larger value of $Q_{\rm Lag}$ than the FZH04 model, by producing a larger number of large bubbles. 
The trends shown by the scale free ICs indicate that an extension of our algorithm to CDM ICs (which are more computationally intensive) should also show only a minor discrepancy between $Q_{\rm Lag}$ and $\zeta f_{\rm src}$.

Our algorithm is by no means a complete solution of the problem, however. For example, our extension to scale free  power spectra is, strictly speaking, inconsistent because it violates the main requirement of statistical independence of the regions being considered \citep[which is guaranteed only for white noise ICs; for further discussion of this point, see][]{sl99b}. Further, as we pointed out in section~\ref{sec:numconsbub}, we continue to use the excursion set ansatz which equates the results of a large number of simulations tracking a single bubble each, with counting many bubbles in a single box. This is almost certainly why our model predicts $Q_{\rm Lag}$ to be slightly smaller than $\zeta f_{\rm src}$ \emph{even for white noise ICs}.

Despite these shortcomings, we would argue that our results present an important step towards building fully self-consistent models of bubble growth. Although, ideally, one would like to build a fully analytical model, perhaps including improvements such as those based on peaks theory \citep{pc14}, our results indicate that this will be quite challenging and one will have to settle with at least a Monte Carlo based approach at some stage of the calculation. Simultaneously, we have seen that current semi-numerical schemes of bubble growth \emph{also} violate photon number conservation (Appendix~\ref{app:seminum}). It therefore appears that the largest returns would arise from developing semi-numerical models that explicitly conserve photon number. We will pursue this line of reasoning in future work.

\section*{Acknowledgments}
We thank Andrei Mesinger for useful comments on the manuscript.
AP gratefully acknowledges the use of facilities at ETH, Z\"urich during the early phases of this work. The research of HP is supported by the Shyama Prasad Mukherjee Fellowship of CSIR, India.

\bibliography{masterRef,bubbleRefs}

\appendix

\section{$\zeta f_{\rm src}$ and $Q_{\rm Lag}$ for the FZH model}
\label{app:zetafcoll}
\noindent
In the standard excursion set formalism, the variation of overdensity in a region as a function of the size is formulated in terms of random walks in the $(\delta, S \equiv \sigma^2)$ plane. In particular, the mass fraction in ionizing sources (i.e., collapsed haloes with $m_h > m_{\rm min}$) is given by
\be
f_{\rm src} = \int_0^{s_{\rm min}} \der S~f(\delta_c, S),
\ee
where $f(\delta_c, S)~\der S$ is the fraction of random walks which first-cross the halo barrier $\delta_c$ between $S$ and $S + \der S$. We have omitted the $z$-dependence of $\delta_c$ for simplicity. Note that we have not made any assumption regarding the specific form of the filter which has been used while calculating the mass variance $\sigma^2(m)$.

In the formalism of FZH04, the bubble barrier $B_{\HII}(S)$ is defined such that the source mass fraction $f(>m_{\rm min}|\delta_0 = B_{\HII}(S), S) = \zeta^{-1}$, and the barrier always satisfies the condition $B_{\HII}(S) < \delta_c$. The fraction of walks which first-cross the bubble barrier between $S$ and $S + \der S$ is denoted as $f_{\HII}(S)~\der S \equiv f(B_{\HII}(S), S)~\der S$. The ionized mass fraction is given by
\be
Q_{\rm Lag} = \int_0^{s_{\zeta {\rm min}}} \der S~f(B_{\HII}(S), S),
\ee
where $s_{\zeta {\rm min}} = \sigma^2(\zeta m_{\rm min})$. 

The bubble barrier is defined only for $S < s_{\zeta {\rm min}}$, which follows from the fact that the minimum size of an ionized bubble should be $\zeta m_{\rm min}$. It is instructive, however, to consider a purely algebraic (and unphysical) extension of $B_{\HII}(S)$ beyond $s_{\zeta {\rm min}}$ such that it lies below $\delta_c$ as long as $S < s_{\rm min}$. The functional form of the FZH04 barrier \eqref{BHII-FZH}, when extended up to $S = s_{\rm min}$, satisfies this condition. The first crossing distribution $f(B_{\HII}(S), S)$ of this barrier then also extends to $S = s_{\rm min}$.

We can then write the mass fraction in sources as
\begin{align}
f_{\rm src} &= \int_0^{s_{\rm min}} \der S~\int_0^S \der S_1~f(\delta_c, S|B_{\HII}(S_1), S_1)\notag\\
&\phantom{\int_0^{s_{\rm min}} \der S~\int_0^S \der S_1}
\times f(B_{\HII}(S_1), S_1),
\end{align}
where $f(\delta_c, S|B_{\HII}(S_1), S_1)~\der S$ is the fraction of walks which first cross the halo barrier between $S$ and $S + \der S$ having first crossed the bubble barrier at $S_1 \leq S$. Interchanging the order of integration gives
\begin{align}
f_{\rm src} &= \int_0^{s_{\rm min}} \der S_1~ f(B_{\HII}(S_1), S_1)\notag\\ 
&\phantom{\int_0^{s_{\rm min}} \der S_1}
\times\int_{S_1}^{s_{\rm min}} \der S~f(\delta_c, S|B_{\HII}(S_1), S_1),
\end{align}
where care must be taken in determining the limits of the integrals. The integral over $S$ is $f(>m_{\rm min} | B_{\HII}(S_1), S_1)$, the mass fraction in sources in a region of scale $S_1$ which is the largest region to have attained an overdensity $B_{\HII}(S_1)$. According to the definition of the bubble barrier, this is simply equal to $\zeta^{-1}$. Hence, the above equation  reduces to
\be
\zeta f_{\rm src} = \int_0^{s_{\rm min}} \der S_1~ f(B_{\HII}(S_1), S_1),
\ee
which is identical to the expression for $Q_{\rm Lag}$ except for the fact that the upper limit of the above integral is $s_{\rm min}$ instead of $s_{\zeta {\rm min}}$\footnote{We have also explicitly checked this by comparing the values of $Q_{\rm Lag}$ returned by Monte Carlo solutions of the first crossing, by sharp-$k$ walks, of the barrier \eqref{BHII-FZH} extended to $S=s_{\rm min}$, for various redshifts and $\zeta$; we indeed find that $Q_{\rm Lag}\approx\zeta f_{\rm src}$ to within statistical errors in all cases.}. 
For $\zeta > 1$, we have $s_{\rm min} > s_{\zeta {\rm min}}$, hence $\zeta f_{\rm src} > Q_{\rm Lag}$. 

Algebraically, therefore, the Lagrangian ionized mass fraction will always be less that the quantity $\zeta f_{\rm src}$ in the standard excursion set formalism for computing the bubble distribution.
The physical reason why these quantities are unequal in this formalism was discussed in the main text. The calculation above emphasizes that it is important to keep track of the limits of various integrations during the calculation, ignoring which can lead to the erroneous result that $Q_{\rm Lag}$ and $\zeta f_{\rm src}$ are represented by the same integral and are hence equal \citep[see, e.g.,][]{zahn+07}. We emphasize that our discussion above \emph{does not depend} on the choice of filter: \emph{any} traditional excursion set model such as that of FZH04 will violate photon number conservation, \emph{regardless} of choice of filter.

\section{Non-conservation of photons in semi-numerical simulations}
\label{app:seminum}
\noindent
Almost all excursion set based semi-numerical simulations of ionized bubbles are based on the formalism of FZH04. Given that the FZH04 explicitly violates the photon number conservation \citep[this was already pointed out by FZH04, and discussed further by][]{zahn+07,zahn+11}, it is worth exploring whether such violation exists in these simulations too. For definiteness, we shall base our discussions on the publicly available semi-numerical code \textsc{21cmFAST}\footnote{http://homepage.sns.it/mesinger/DexM\texttt{\_\_\_}21cmFAST.html} \citep{mfc11-21cmfast}, although most of the conclusions would hold for other such codes as well \citep[e.g.,][]{chr09}. 

\begin{figure}
\centering
\includegraphics[width=0.45\textwidth]{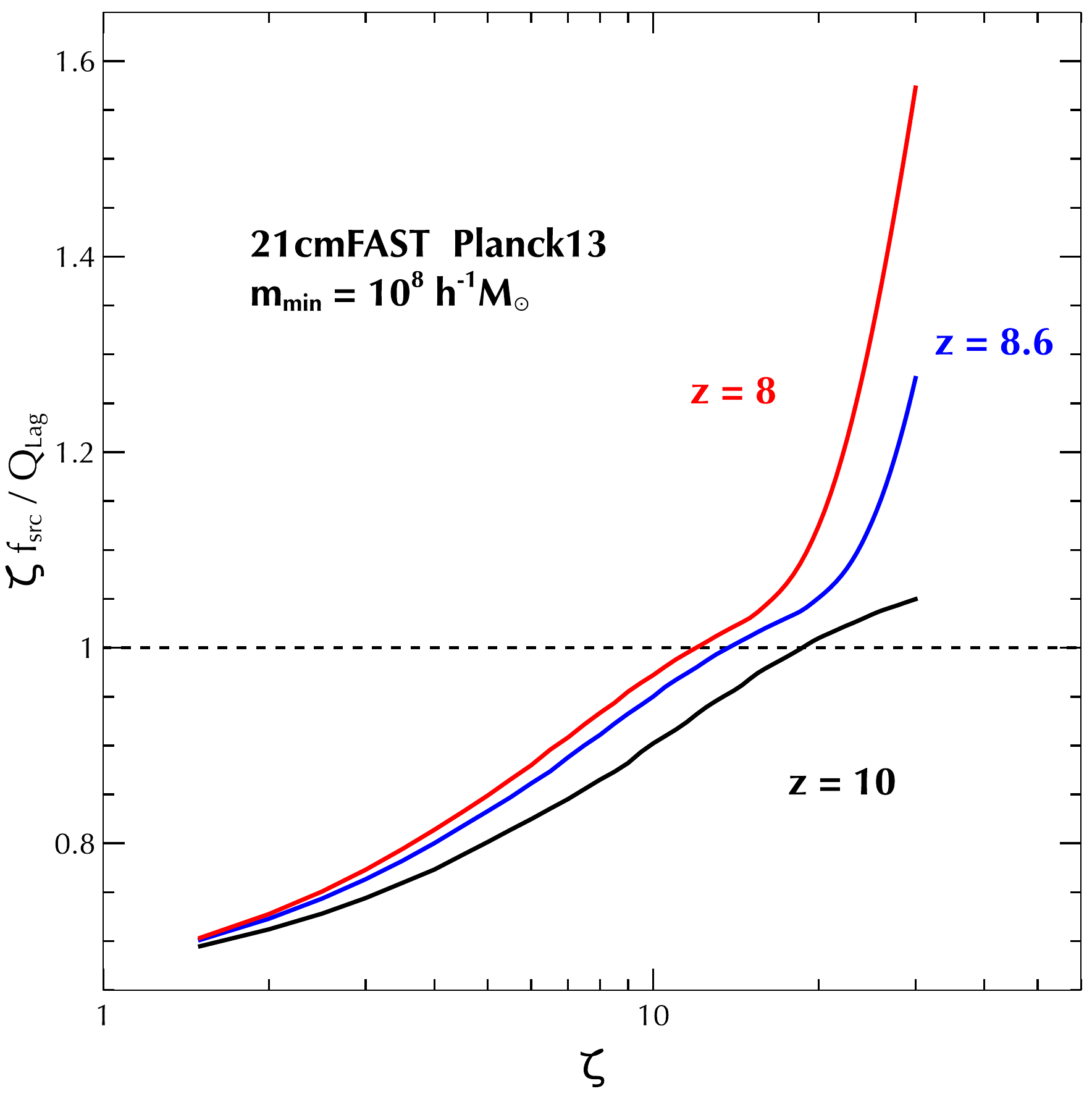}
\caption{Ratio of $\zeta f_{\rm src}$ and $Q_{\rm Lag}$ as inferred from the output of the \textsc{21cmFAST} code, for multiple redshifts as a function of $\zeta$, assuming $\Lambda$CDM initial conditions with a Planck13 cosmology and $m_{\rm min}=10^8\Mh$.}
\label{fig:QLagVszetafc-21cmFast}
\end{figure}

We show in Figure~\ref{fig:QLagVszetafc-21cmFast}  the ratio $\zeta f_{\rm src} / Q_{\rm Lag}$ for a default run\footnote{The default settings use a (400 Mpc)$^3$ comoving box with a $200^3$ grid, giving a spatial resolution of 2 Mpc comoving, and use the sharp-$k$ filter to identify ionized regions.} of \textsc{21cmFAST}, with cosmological parameters and $m_{\rm min}$ set to the same values as in the left panel of Figure~\ref{fig:QLagVszetafc-FZHfull}. We also ensured that the quantity $Q_{\rm Lag}$ is interpreted to be the mass-averaged ionized fraction (as opposed to the volume-averaged quantity). One can immediately see from the plot that \textsc{21cmFAST} too violates photon number conservation, and the violation can be as large as $40$ per cent either way. The other point to note is that even though \textsc{21cmFAST} is based on the FZH04 formalism, the results obtained from the two do not agree with each other. This is not surprising because there exist subtle differences in the approaches, e.g., (i) the FZH04 model computes the collapsed fraction of ionizing sources using the excursion set model of \citet{bcek91}, while \textsc{21cmFAST} scales the collapsed fraction to match the globally averaged collapsed fraction as given by the Sheth-Tormen formula \citep{st99,jenkins+01}, and (ii) \textsc{21cmFAST}, while identifying ionized cells using spheres of varying size, flags only the central cell in a spherical region as ionized, thus essentially disregarding any information on the particular scale at which the ionization barrier is crossed.

However, unlike FZH04 where we always have $Q_{\rm Lag} < \zeta f_{\rm src}$, the \textsc{21cmFAST} output gives $Q_{\rm Lag} > \zeta f_{\rm src}$ for small values of $\zeta$. This was noted earlier by \citet{zahn+07} who, using a toy model, showed that the value of $Q_{\rm Lag}$ can be larger or smaller than $\zeta f_{\rm src}$ depending on the distance between the sources. The discussion in \citet{zahn+07} seems to imply that such a discrepancy is mainly due to the use of top-hat filtering in real space and should not be present if bubbles are identified using sharp-$k$ filtering. However, the default implementation of \textsc{21cmFAST} does use sharp-$k$ filtering, so this argument does not explain the results of Figure~\ref{fig:QLagVszetafc-21cmFast}. In fact, while the precise origin of photon non-conservation in semi-numerical models is unclear, our analytical results in Appendix \ref{app:zetafcoll} suggest that this would be independent of the choice of filter.

The non-conservation of photon number is not limited to \textsc{21cmFAST} and is seen in other types of semi-numerical codes too, e.g., codes where the density field and collapsed haloes are generated using a full $N$-body simulation \citep{zahn+07,chr09}.  Many of these semi-numerical calculations have been found to match the results of full radiative transfer simulations which are expected to account for the photon numbers self-consistently without violating any conservation \citep{zahn+07,mfc11-21cmfast,majumdar+14}, hence one might ask why the photon non-conservation does not show up in these comparison studies. The reason for this is that the comparisons are usually carried out for a fixed value of the ionized fraction (either volume or mass weighted). In other words, the value of $\zeta$ in the semi-numerical calculations is adjusted to obtain the value of ionized fraction as given by the radiative transfer simulations, and the resulting distributions compared thereafter. Since the morphologies of the ionized regions obtained from the two methods are quite similar at a fixed ionized fraction $Q_{\rm Lag}$, the non-conservation of photons which impacts the relation between $Q_{\rm Lag}$ and $\zeta$ can be calibrated away for the semi-numerical simulations.  It is thus not surprising the issue of non-conservation did not show up in those studies (see, e.g., the right panel of Figure~\ref{fig:sizedist} in the main text which shows a comparison between the FZH04 result and our approximately photon-conserving algorithm). It would be interesting to explore the reasons for the non-conservation in semi-numerical models and possibly work out a corresponding photon-conserving algorithm. 

\section{The partition algorithm of SL99}
\label{app:SL99}
\noindent
In this section we describe the algorithm of SL99 which allows us to partition a given region of mass $M_0$ and overdensity $\delo$ into (sub-)haloes. The basic idea of the method is to start by choosing the mass of the very first halo, then choose the mass of a second one from the mass (and volume) that remains in the region and keep continuing the process until all the mass $M_0$ has been assigned to halos. In practice, one would stop when the remaining available mass in the region falls below some chosen threshold. Our application to creating ionized bubbles gives us a natural value for this threshold, namely, the minimum source mass $m_{\rm min}$.

We know that the fraction of mass contained within haloes of mass $m$ within the region under consideration is given by \citep{bcek91}
\be
f(m|\delo, M_0)\der m = \frac{\delta_c - \delo}{\sqrt{2 \pi} (s - S_0)^{3/2}} {\rm e}^{-(\delta_c - \delo)^2/2 (s - S_0)}\der s,
\ee
where $s = \sigma^2(m)$ and $S_0 = \sigma^2(M_0)$. Hence, one should choose the mass of the halo according to the probability distribution given by $f(m|\delo, M_0)$. This can be done by drawing a standard Gaussian random deviate $\nu$, setting $s = S_0 + (\delc-\delo)^2/\nu^2$ and then inverting the relation $s=\sigma^2(m)$.

Assuming the first halo has a mass $m_1$, the mass remaining in the region is simply $M'= M_0 - m_1$. The overdensity $\delta'$ in the remaining volume can be obtained from the expression
\be
\delta_c - \delta' = \frac{\delta_c - \delo}{1 - m_1/M_0}
\ee
where we have kept the leading order terms in \delo\ and $\delta'$.

One can then apply the same algorithm on the remaining mass to further partition it, replacing $(\delo,M_0)$ with $(\delta',M')$, \emph{provided this region is statistically independent of the mass in the already partitioned region}. This condition is satisfied only when the initial conditions have a white noise power spectrum. Assuming that is the case, one simply needs to repeat the algorithm updating the mass and density of the remaining region. If required, one can also run the whole partitioning algorithm starting with different random seeds and construct random realizations of the halo distribution in the given $(\delo, M_0)$ region. The method, by construction, guarantees that the halo distribution generated has mean values which are identical to that obtained by the excursion set formalism.

\label{lastpage}

\end{document}